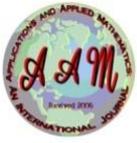

Available at
http://pvamu.edu/aam
Appl. Appl. Math.
ISSN: 1932-9466

**Applications and Applied Mathematics:**
An International Journal
**(AAM)**

Vol. 14, Issue 2 (December 2019), pp. 777 - 783# A further result on the aging properties of an extended additive hazard model

[1]Morteza Raeisi and [2*]Gholamhossein Yari

[1]Department of Statistics
Science and Research Branch
Islamic Azad University
Daneshgah Blvd., Simon Bulivar Blvd.
Tehran, Iran
mreisie@gmail.com;

[2]School of Mathematics
Iran University of Science and Technology
Hengam St., Resalat Square
Tehran, Iran
yari@iust.ac.ir;

* Corresponding authorReceived: August 18, 2018; Accepted: November 11, 2019## Abstract

The passing of time is an important factor for covariates in the additive and proportional hazard models. According to this idea, the extended additive hazard model (EAHM) is introduced by considering the time-varying effects of covariates and is investigated several properties of this model related to reliability analysis. In this paper, we obtain a further result for the EAHM with respect to the aging properties.

**Keywords:** Additive hazard model; stochastic orderings; Aging properties; Reliability; Extended additive hazard model; Totally positive (reverse regular)

**MSC 2010 No.:** 62N05; 60K10## 1. Introduction

One of the main purposes of survival analysis is to investigate the effects of risk factors on disease or death occurrence. When the effect of a factor under study has a multiplicative effect on the

777

baseline hazard function, we have a proportional hazard model and an additive effect leads to an additive hazard model. The proportional hazard model is mostly studied in the literature. However, recently, study the additive hazard model be more in terms of quantity in the survival analysis literature (e.g. Li and Ling (2012); Nair and Sankaran (2012); Zhang and Zhang (2014); Gupta (2016); Kayid et al. (2016); Esna–Ashari and Asadi (2016); Raeisi and Yari (2018); Liu et al. (2018) and Kotelo (2019)) because from the public health point of view, to study the risk difference is more important than the risk ratio in describing the association between the risk factor and the occurrence of a disease. Thus, the additive models have been studied by different researchers that making the additive models more important (Das and Nanda (2013)).

The additive hazard model has been initiated by Aalen (1980) and Aalen (1989). Nair and Sankaran (2012) suggested a more flexible form and considered that if $X^*$ denotes the lifetime variable in population with hazard rate function $h^*$, then the conditional hazard rate function of $X^*|Z = z$ is given for all x ≥ 0 by

$$h^*(x|z) = a(z) + h(x), \tag{1}$$

for some positive function $a(z)$. They have studied the analytical properties of the additive hazard model and have compared the aging properties of the baseline random variable and the induced random variable. Li and Ling (2012) have discussed the aging and dependence properties in the additive hazard mixing model. Some stochastic comparisons have also been studied in this paper. The identity of (1) aims to model the conditional hazard rate of $X^*$ given that $Z = z$ is observed for the covariate variable, using an additive effect function of z, on the baseline hazard function. In survival analysis and reliability, systems (which can be a machine, human, ...) operate in a changing environment. The conditions under which systems operate can be harsher or gentler in modeling the lifetime of the systems and these conditions can be also changed in passing of the time. Hence, Raeisi and Yari (2018) introduced an extended additive hazard model (EAHM) for all x ≥ 0, as following

$$h^*(x|z) = a(z, x) + h(x), \tag{2}$$

for some positive function $a(z, x)$ for all x ≥ 0, satisfying $\int_0^\infty a(x, z)dz = \infty$, for each z ≥ 0. They investigated the model in two parts. First, they introduced the extended model and concentrated on the distribution theory of the EAHM by giving representations of the model via some reliability measure and by a number of useful illustrative examples. Then, they investigated some further results for additive hazard model (1) proposed by Nair and Sankaran (2012) and several closure properties of the EAHM (2) with respect to some dependence structures, stochastic orderings, and aging properties and obtained also some preservation properties of the EAHM under some conditional stochastic orderings. Indeed, in EAHM, the additive effect on baseline hazard function is considered as a function of both x and z because the susceptibility to causes of failure at any given amount of the observed value of the covariate random variable $Z$ is different over the time. By this approach, heterogeneity of susceptibility to causes of failure of a device at any observed values of covariate $Z$ over the time during, which the device is working, is well-explained in the additive hazard model.

In this paper, in Section 2, we present some definitions and basic properties that we will use through the paper. In Section 3, we investigate the closure properties of the model with respect to the aging properties. In Section 4, we discuss some conclusions.

## 2. Preliminaries

In this section, for ease of reference, we present some definitions and basic properties that be used through the paper. First, we present definitions of some stochastic orders and aging notions. For the stochastic orders, we refer to Shaked and Shanthikumar (2007) and Belzunce et al. (2015). For the aging notions, we refer to Lai and Xie (2006) and Barlow and Proschan (1981). Through this paper X and Y are two nonnegative random variables with distribution functions F and G, survival functions $\bar{F} = 1 - F$ and $\bar{G} = 1 - G$, density functions $f$ and g, and hazard rate functions $h_X(x) = f(x)/\bar{F}(x)$ and $h_Y(x) = g(x)/\bar{G}(x)$, respectively.

**Definition 2.1.**

A random variable X is said to be smaller than Y in the:

(i) Likelihood ratio order (denoted as $X \leq_{LR} Y$), if $g(x)/f(x)$ is increasing in $x \geq 0$.
(ii) Usual stochastic order (denoted as $X \leq_{ST} Y$), if $F(x) \leq G(x)$, for all $x \geq 0$.

**Definition 2.2.**

The nonnegative random variable X is said to have:

(i) Increasing (decreasing) likelihood ratio (ILR(DLR)) property, if $f$ is a log-concave (log-convex) function on $(0, \infty)$.
(ii) Increasing (decreasing) Failure rate (IFR(DFR)) property, if $h_X(x)$ is increasing (decreasing) in $x \geq 0$, or equivalently if $\bar{F}(x)$ is log-concave (log-convex) for $x \geq 0$.

Karlin (1968) introduced the concept of sign-regular of order 2, which is of great importance in the fields of mathematics and statistics with various applications. The definition of sign-regular of order 2 leads to a particular case as follows.

**Definition 2.3.**

A nonnegative function $\beta(x, y)$ is said to be totally positive (reverse regular) of order 2, denoted as $TP_2$ ($RR_2$), in $(x, y) \in \chi \times \gamma$, if

$$\beta(x_1, y_1)\beta(x_2, y_2) - \beta(x_1, y_2)\beta(x_2, y_1) \geq (\leq) 0,$$

for all $x_1 \leq x_2 \in \chi$ and $y_1 \leq y_2 \in \gamma$ in which $\chi$ and $\gamma$ are two subsets of the real line $\mathbb{R}$.

## 3. Main Results

Raeisi and Yari (2018) investigated the extension and more flexibility of EAHM (2) over AHM (1). Moreover, one can show the Theorems 4.5 and 4.6 in Raeisi and Yari (2018) for proof just change $\exp(-w(x,z))$ by $\exp(-xa(z))$ and this is another reason for extending (1) by (2).
As Nair and Sankaran (2012) pointed out, comparison of the different ageing properties of X and $X^*$ will often help in the selection of appropriate model and its analysis in a practical situations. In this section, we discuss preservation properties of some aging notions under the transformation $X \to X^*$ in the EAHM (2). Before stating the results, we must state the following useful lemma.

**Lemma 4.1.** (Misra and van der Meulen (2003))

Let $V \geq 0$ be a random variable with distribution function from the family $\mathcal{F}=\{G_\theta, \theta \in \Theta \subseteq \mathbb{R}\}$ such that $G_{\theta_1} \leq_{ST} (\geq_{ST}) G_{\theta_2}$, for all $\theta_1 \leq \theta_2 \in \Theta$. Suppose a measurable real function $\varphi(\theta, v)$ for which $\mathbb{E}_\theta[\varphi(\theta, v)]$ exists. Then, the function $\mathbb{E}_\theta[\varphi(\theta, v)]$ is

(i) Decreasing in $\theta$, if $\varphi(\theta, v)$ is decreasing (increasing) in $v$ and decreasing in $\theta$,
(ii) Increasing in $\theta$, if $\varphi(\theta, v)$ is increasing (decreasing) in $v$ and increasing in $\theta$.

It is well-known in the literature (e.g. see Shaked and Shanthikumar (2007)) that the log-convexity property of the density function implies the log-convexity property of the survival function. In other words, the DLR property is stronger than the DFR property. Therefore, it is useful to study conditions under which the DFR property of the baseline variable is translated to the DLR property of the overall variable in the EAHM.

**Theorem 4.1.**

Let $X$ be DFR, let $a(x, z)$ be decreasing in $x$, for all $z \geq 0$, and decreasing (increasing) in $z$, for any $x \geq 0$, and let $h^*(x \mid z)$ be $TP_2$ ($RR_2$) in $(x, z) \in [0, \infty) \times [0, \infty)$, such that it is log-convex in $x$, for any $z \geq 0$. Then $X^*$ is DLR.

*Proof:*

In view of Raeisi and Yari (2018), the density function of $X^*$ is given by

$$f^*(x) = f(x)\mathbb{E}(e^{-w(x,Z)}) + S(x)\mathbb{E}(a(x,Z)e^{-w(x,Z)})$$
$$= S(x)\mathbb{E}\left((a(x,z) + h(x))e^{-w(x,Z)}\right)$$
$$= S(x)\mathbb{E}(h^*(x \mid Z)e^{-w(x,Z)}),$$

for all $x \geq 0$, where $w(x, z) = \int_0^x a(t, z)dt$ and $S(x)$ is survival function of X. To get the proof it suffices to prove that $f^*(x + \theta)/f^*(\theta)$ is increasing in $\theta$, for any $x \geq 0$. First, we can see that

$$\frac{f^*(x+\theta)}{f^*(\theta)} = \frac{S(x+\theta)}{S(x)} \frac{\mathbb{E}(h^*(x+\theta \mid Z)e^{-w(x+\theta,Z)})}{\mathbb{E}(h^*(x \mid Z)e^{-w(x,Z)})}$$

$$= \frac{S(x+\theta)}{S(x)} \int_0^\infty e^{w(\theta,v)-w(x+\theta,v)} \frac{h^*(x+\theta|v)}{h^*(\theta|v)} \frac{h^*(\theta|v)g(v)}{\int_0^\infty h^*(\theta|v)g(v)dv} dv$$

$$= \frac{S(x+\theta)}{S(x)} \mathbb{E}(\varphi(\theta,V)),$$

where $\varphi(\theta,v) = e^{w(\theta,v)-w(x+\theta,v)} \frac{h^*(x+\theta|v)}{h^*(\theta|v)}$, for all $x, \theta \geq 0$ and $V$ is a nonnegative random variable with density function $g_\theta$, for any $\theta \geq 0$, given by

$$g_\theta(v) = \frac{h^*(\theta|v)g(v)}{\int_0^\infty h^*(\theta|v)g(v)dv}, \quad v \geq 0.$$

When $X$ is DFR, the ratio $S(x+\theta)/S(\theta)$ is increasing in $\theta$. Thus, it is enough to show by Lemma 4.1 that $\mathbb{E}(\varphi(\theta,V))$ is increasing in $\theta$. Let us observe first that

$$\begin{aligned} w(\theta,z) - w(x+\theta,z) &= \int_0^\theta a(t,z)dt - \int_0^{x+\theta} a(t,z)dt \\ &= -\int_\theta^{x+\theta} a(t,z)dt \\ &= -\int_0^x a(t+\theta,z)dt, \end{aligned}$$

for all $x, \theta, z \geq 0$. Because, by assumption, $a(x,z)$ is decreasing in x, for all $z \geq 0$, it is therefore clear from above that $w(\theta,z) - w(x+\theta,z)$ is increasing in $\theta$, for all $x, z \geq 0$. In addition, since $h^*(x|z)$ is, from assumption, log-convex in x, for any $z \geq 0$, thus $h^*(x+\theta|z)/h^*(\theta|z)$ is increasing in $\theta$, for each $z \geq 0$. It now follows that $\varphi(\theta,v)$ is increasing in $\theta$, for all $v \geq 0$. On the other hand, we can observe that if $a(x,z)$ is decreasing (increasing) in z, for all $x \geq 0$ and $h^*(\theta|z)$ is $TP_2(RR_2)$ in $(x,z) \in [0,\infty) \times [0,\infty)$, then $\varphi(\theta,v)$ is increasing (decreasing) in $v \geq 0$, for all $\theta \geq 0$. Denote by $G_\theta$ the distribution function of V, for any $\theta \geq 0$. Then, it is easy to see that $h^*(\theta|z)$ is $TP_2(RR_2)$ in $(x,z) \in [0,\infty) \times [0,\infty)$, if and only if that $G_{\theta_1} \leq_{LR} (\geq_{LR}) G_{\theta_2}$, for all $\theta_1 \leq \theta_2 \in [0,\infty)$. Because the LR order implies the ST order (cf. Shaked and Shanthikumar (2007)), it consequently follows that that $G_{\theta_1} \leq_{ST} (\geq_{ST}) G_{\theta_2}$, for all $\theta_1 \leq \theta_2 \in [0,\infty)$. An application of Lemma 4.1 (ii) completes the proof. ∎

## 4. Conclusion

In this paper, we obtained one property of the aging notion under the transformation $X \to X^*$ in the EAHM. Similar results can be obtained in the case of the aging concepts, DMRLHA, NBUC, UBA, and UBAE (see Lai and Xie (2006) and Barlow and Proschan (1981) for the definition of these concepts) using the same kind of arguments.

The extended additive hazard model can be used in medical and epidemiological studies. For example, since it can be possible that a new and better drug is found after passing the duration of time, it implies that passing of the time has an impact on covariates (e.g. sex, age, the effect of drugs and …). Thus, the hazard rate can change that this change should be considered in the model. We expect that the proposed approach by ours gives better results that should investigate it in real data as future work.

*Acknowledgment*

*The authors are thankful to anonymous referees and the Editor-in-Chief, Professor Aliakbar Montazer Haghighi for useful comments and suggestions to improve the paper.*

## Authors biographical

[1]Morteza Raeisi was born in 1984 in Iran. He is a PhD student of statistics at Science and Research Branch Islamic Azad University, Tehran, Iran. His interests are survival analysis and mixture models.

[2]Gholamhossein Yari was born in 1954 in Iran. He is an associated professor of statistics at Iran University of Science and Technology, Tehran. His interests are information theory and stochastics processes.